\documentclass{iau307}
\usepackage{graphicx}
\usepackage{natbib}
\usepackage{url}
\usepackage{dtklogos}
\bibpunct{(}{)}{;}{a}{}{,}

\newcommand{\msun}{\ensuremath{M_{\odot}}}
\newcommand{\lsun}{\ensuremath{L_{\odot}}}
\newcommand{\rsun}{\ensuremath{R_{\odot}}}

\newcommand{\Teff}{\ensuremath{T_{\rm eff}}}
\newcommand{\vinf}{\ensuremath{v_{\infty}}}
\newcommand{\mdot}{\ensuremath{\dot{M}}}

\newcommand{\msunyr}{\ensuremath{M_{\odot} {\rm yr}^{-1}}}

\newcommand{\mdu}{\ensuremath{10^{-6}\,M_{\odot} {\rm yr}^{-1}}}

\newcommand{\beq}{\begin{equation}}
\newcommand{\eeq}{\end{equation}}
\newcommand{\beqa}{\begin{eqnarray}}
\newcommand{\eeqa}{\end{eqnarray}}

\newcommand{\nbeq}{\begin{equation*}}
\newcommand{\neeq}{\end{equation*}}

\newcommand{\kms}{\ensuremath{{\rm km}\,{\rm s}^{-1}}}

\newcommand{\rarrow}{\rightarrow}
\newcommand{\dd}{{\rm d}}

\newcommand{\Ha} {H$_{\rm \alpha}$}

\newcommand{\Rstar}{\ensuremath{R_{\ast}}}

\newcommand{\Lstar}{\ensuremath{L_{\ast}}}

\newcommand{\Dmom}{\ensuremath{D_{\rm mom}}}

\newcommand{\vesc}{\ensuremath{v_{\rm esc}}}

\newcommand{\dvdr}{\ensuremath{\dd v/\dd r}}

\newcommand{\grad}{\ensuremath{g_{\rm rad}}}

\title[Physics of mass loss in massive stars] %% give here short title %%
{Physics of Mass Loss in Massive Stars}

\author[Puls, Sundqvist, \& Markova]   %% give here short author list %%
{Joachim Puls$^1$
 \and Jon O. Sundqvist$^1$
 \and Nevena Markova$^2$}

\affiliation{$^1$Universit\"atssternwarte, Scheinerstr. 1, D-81679
M\"unchen; Germany \\ email: {\tt uh101aw@usm.uni-muenchen.de} \\[\affilskip]
$^2$Institute of Astronomy with NAO, BAS, PO Box 136, 4700 Smolyan, Bulgaria}

\pubyear{2014}
\volume{307} 
\pagerange{}
\jname{New windows on massive stars: asteroseismology, interferometry, and spectropolarimetry}
\editors{G. Meynet, C. Georgy, J.H. Groh \& Ph. Stee, eds.}
\begin{document}

\maketitle

\begin{abstract}
We review potential mass-loss mechanisms in the various
evolutionary stages of massive stars, from the well-known line-driven
winds of O-stars and BA-supergiants to the less-understood winds from
Red Supergiants. We discuss optically thick winds from
Wolf-Rayet stars and Very Massive Stars, and the hypothesis of
porosity-moderated, continuum-driven mass loss from stars formally
exceeding the Eddington limit, which might explain the giant outbursts
from Luminous Blue Variables. We finish this review with a glance on
the impact of rapid rotation, magnetic fields and small-scale
inhomogeneities in line-driven winds.

\end{abstract}

\firstsection % if your document starts with a section,
\section{Introduction} 

Stellar winds from massive stars are fundamentally important, in
providing energy and momentum input into the ISM, in creating
wind-blown bubbles and circumstellar shells, and in triggering star
formation. The presence and amount of mass loss decisively controls
the evolution and fate of massive stars\footnote{A change of only a
factor of two in the mass-loss rates can have a dramatic effect
\citep{meynet94}.} (e.g., the type of final Supernova-explosion), by
modifying evolutionary timescales, surface abundances, and stellar
luminosities. Moreover, mass loss also affects the atmospheric structure,
and only by a proper modeling of stellar winds it is possible to
derive accurate stellar parameters by means of quantitative
spectroscopy. In the following, we review the status quo of our
knowledge about the physics of these winds.

\section{Basic considerations\label{basics}} 

Since massive stars have a high luminosity, they are basically able to
generate a large radiative acceleration in their atmospheres. 

\smallskip
\noindent
{\bf Global energy budget -- The photon tiring limit.} Thus, a first
question regards the maximum mass-loss rate that can be radiatively
accelerated. By equating the mechanical luminosity in the wind at
infinity with the available photospheric luminosity \Lstar\ (in this case, all
photons have used up their energy/momentum, $L(\infty)=0$), one
obtains
\beq
\mdot_{\rm max}=\frac{2 \Lstar}{\vinf^2+\vesc^2} = \frac{\mdot_{\rm
tir}}{1+(\vinf/\vesc)^2} 
\approx \mdot_{\rm tir} {\ldots} \frac{\mdot_{\rm tir}}{10},
\eeq
where \vinf\ is the terminal wind speed (typically on the order of one to three
times the photospheric escape velocity, $\vesc = \sqrt{2GM/R}$), and $\mdot_{\rm tir}$
the `photon tiring mass-loss rate' \citep{OG97}, i.e., the maximum
mass-loss rate when the wind just escapes the gravitational potential, 
with $\vinf \rightarrow 0$. In convenient units,
\beq
\mdot_{\rm tir}=\frac{2 \Lstar}{\vesc^2} =
0.032\frac{\msun}{yr}\frac{\Lstar}{10^6 \lsun}\frac{R/\rsun}{M/\msun}.
\eeq
This maximum mass-loss rate is much larger than the mass-loss rates of
winds from OB-stars/A-supergiants/WR-stars/Red Supergiants(RSGs) (by a factor of 
$10^3$ and larger), whereas it is on the order of the mass-loss rates
estimated for the giant eruptions from Luminous Blue Variables (LBVs,
see Sect.~\ref{porwinds}).

\smallskip
\noindent
{\bf Global momentum budget -- optically thick/thin winds.}
The dominating terms governing the equation of motion of a
massive star wind are the inward directed gravitational pull and the outward
directed radiative acceleration, \grad\footnote{as long as pressure terms can
be neglected, i.e., when $v(r) \gg v_{\rm sound}$ for the largest part
of the wind.}. In spherical symmetry, the latter can be written as
\beq
\label{gradformal}
\grad(r)=\int_0^\infty d\nu \frac{\kappa_\nu(r) F_\nu(r)}{c} = \kappa_{\rm F}(r)
\frac{\Lstar}{4\pi r^2 c},
\eeq
with frequential flux $F_\nu$, mass absorption coefficient
$\kappa_\nu$, and flux-weighted mass absorption coefficient
$\kappa_{\rm F}$. By integrating the equation of motion over $dm =
4\pi r^2\rho dr$ between the sonic point, $r_{\rm s}$, and infinity (and
neglecting pressure terms), one can express the wind-momentum rate,
$\mdot \vinf$, in terms of the total momentum rate of the radiation
field, $\Lstar/c$, and flux mean optical depth of the wind, $\tau_{\rm
F}$,
\beq
\eta = \frac{\mdot \vinf}{\Lstar/c} \approx \tau_{\rm F}(r_{\rm s}) - \frac{\tau_{\rm e}(r_{\rm s})}{\Gamma_{\rm e}}
\eeq
(cf. \citealt{abbott80}). The second term on the rhs is a typically
small correction for overcoming the gravitational potential,
consisting of the electron-scattering optical depth $\tau_{\rm e}$, and the
conventional Eddington-Gamma, $\Gamma_{\rm e} \propto L/M$, evaluated for
electron-scattering. Note that this relation is only valid for $\mdot
\ll \mdot_{\rm tir}$. 

Because of its definition, $\eta$ is called the wind performance
number. For optically thin winds, defined by $\tau_{\rm F}(r_{\rm s}) < 1$, the
performance number is lower than unity, $\eta < 1$, which is typical
for OB-stars and A-supergiants, whilst for optically thick winds
(e.g., WR-winds) with $\tau_{\rm F}(r_{\rm s}) >1$ also $\eta > 1$. In a
line-driven wind (see Sect.~\ref{ldw}), $\eta$ becomes roughly unity when
each photon in the wind is scattered once. $\eta > 1$ then
indicates that most photons have been scattered more than once (multi-line
scattering).

\section{(Optically thin) Line-driven winds\label{ldw}} 

The winds from OB-stars and A-supergiants, with typical mass-loss rates
$\mdot \approx 10^{-7} {\ldots} 10^{-5}$ \msun/yr, and terminal
velocities, \vinf, ranging from 200 {\ldots} 3,500 \kms, are thought
to be accelerated by radiative line-driving. 

Photospheric light is scattered/absorbed in spectral lines, and
momentum is transferred to the absorbing ions, predominantly
into the radial direction. Note that there is no momentum loss or gain
during the (re-)emission process, at least in a spherically symmetric
configuration, since this process is fore-aft symmetric. Most of the
momentum-transfer is accomplished via metallic resonance lines, and
this momentum is then further transferred from the accelerated
metal-ions (with a low mass-fraction) to the wind bulk plasma,
H and He, via Coulomb collisions (e.g., \citealt{SP92}).  

Since the complete process requires a large number of photons (i.e., a
high luminosity), such winds occur in the hottest stars, like O-type
stars of all luminosity classes, but also in cooler BA-supergiants,
because of their larger radii. Efficient line-driving further requires
a large number of spectral lines close to the flux-maximum and a high
interaction probability (i.e., a significant line optical depth). Since
most spectral lines originate from various metals, a strong dependence
of \mdot\ on metallicity is thus to be expected, and such {\it
line-driven} winds should play a minor role (if at all) in the early
Universe (but see Sect.~\ref{porwinds}).

The theory of line-driven winds has been pioneered by
\citet{LS70} and particularly by \citet[`CAK']{CAK}, with
essential improvements regarding a quantitative description and
application provided by \citet{FA86} and \citet{PPK}. Line-driven
winds have been reviewed by \citet{KP00} and more recently by
\citet{puls08b}. 

In the following, we will briefly consider some
relevant aspects, mostly in terms of the `standard model', assuming a
steady-state, spherically symmetric, and homogeneous outflow (i.e.,
neglecting rotation, magnetic fields, and density inhomogeneities, 
considered later in Sect.~\ref{addeffects}).

Calculating the radiative acceleration  by means of the
Sobolev approximation \citep{Sobo60}, assuming well-separated lines
(justified for most optically thin winds, e.g., \citealt{Puls87}),
and a distribution of line-strengths following a power-law with
exponent $\alpha-2$ (for details, see \citealt{Puls00}), the {\it
total} radiative line acceleration from all participating lines can be
expressed by
\beq
\grad(\mbox{all lines}) \propto \Bigl(\frac{\dvdr}{\rho}\Bigr)^\alpha.
\eeq
Particularly because of the dependence on $\rho$, this leads to a
self-regulation of the mass-loss rate, and an analytic solution of the
equation of motion is possible. In compact notation, and neglecting
the (for O-stars typically weak) effects of an ionization
stratification\footnote{corresponding to a `force-multiplier' parameter
$\delta = 0$, see \citet{abbott82} \hfill \quad}, one finds the
following scaling laws
\beq
\label{mdotscal}
\mdot \approx \frac{\Lstar}{c^2} \frac{\alpha}{1-\alpha} 
\Bigl(\frac{{\bar Q} \Gamma_{\rm e}}{1-\Gamma_{\rm e}}\Bigr)^{1/\alpha-1}
\frac{1}{(1+\alpha)^{1/\alpha}}
\eeq
\citep[and references therein]{Owocki04},
\beq
\label{vinfscal}
v(r)  =  \vinf\Bigl(1-\frac{\Rstar}{r}\Bigr)^\beta \qquad	    
\vinf \approx 2.25 \frac{\alpha}{1-\alpha} \vesc'
\eeq
\citep{Kud89}, where $\vesc' = \vesc \sqrt{1-\Gamma_{\rm e}}$ is the
effective escape velocity corrected for Thomson acceleration. For
typical O-stars, \nocite{Gayley95} Gayley's (1995) dimensionless
line-strength parameter $\bar Q \approx 2000$, $\alpha \approx 0.6$,
$\beta \approx 0.8$, and \mdot\ is on the order of $\mdu \ll
\mdot_{\rm tir}$.  Note that $\bar Q$ scales with metallicity, $\bar Q
\propto Z/Z_{\odot}$, such that $\mdot \propto (Z/Z_{\odot})^{0.7}$ for the
above $\alpha$.

For quantitative results, the most frequently used theoretical
mass-loss rates are based on the wind models by \citet{vink00,
vink01}, calculated by means of approximate NLTE occupation numbers
and a Monte Carlo transport (i.e., without invoking any line
statistics). From interpolating the mass-loss rates derived in this
way for a large model grid, the provided `mass-loss recipe' becomes
$\mdot = \mdot(\Lstar, M, \Teff, \vinf/\vesc, Z)$, with a similar
metallicity dependence as above. For alternative models and
calculation methods, see, e.g., \citet{krticka00, pauldrach01, kud02}.

\smallskip
\noindent
{\bf The wind-momentum luminosity relation (WLR).} By using the
scaling relations for \mdot\, and \vinf\,
(Eqs.~\ref{mdotscal}, \ref{vinfscal}), and approximating $\alpha
\approx 2/3$, one obtains the so-called wind-momentum luminosity
relation -- WLR -- \citep{Kud95},
\beq
\label{wlr}
\log \Dmom = \log \Bigl(\mdot \vinf
\bigl(\frac{R}{\rsun}\bigr)^{\frac{1}{2}}\Bigr) \approx \frac{1}{\alpha}
\log \Bigl(\frac{\Lstar}{\lsun}\Bigr) + \mbox{offset({\it Z}, spectral type)},
\eeq
which relates the {\it modified} wind-momentum rate \Dmom\, with 
only the stellar luminosity. The mass-dependence (due to
$\Gamma_{\rm e}$) becomes negligible as long as $\alpha$ is close to
2/3. The offset in Eq.~\ref{wlr} depends on metallicity and spectral
type, mostly because the effective number of driving lines and thus
$\bar Q$ depend on these quantities (e.g., \citealt{Puls00}), via
different opacities and contributing ions. 

Though derived from simplified scaling relations, the WLR concept has
also been confirmed by numerical model calculations, e.g., those from
\citet[their Fig. 9]{vink00}. 
%Though originally suggested as an independent tool to measure
%extragalactic distances, nowadays the WLR is mostly used to test the
%validity of the line-driven wind theory itself. 
An impressive {\it observational} confirmation of this concept has
been provided by \citet{mokiem07b}, compiling observed stellar and
wind parameters from Galactic, LMC and SMC O-stars, and analyzing the
corresponding WLRs. Accounting for wind-inhomogeneities (see
Sect.~\ref{inhom}) in an approximate way, they derive $\mdot \propto
(Z/Z_{\odot})^{0.72 \pm 0.15}$, in very good agreement with
theoretical predictions. 
%{JS: Still don't like this.. Imho, they are 
%obtaining a result consistent with theory by adopting a clumping-correction 
%initially derived in order to fit the theory (Vink-rates).... But well.}

\section[(Optically thick) Winds from WR- and Very Massive Stars] 
{(Optically thick) Winds from WR- and Very Massive Stars\footnote{see 
also Gr\"afener, this Volume \hfill \quad}}

From early on, the mass-loss rates of Wolf-Rayet stars posed a serious
problem for theoretical explanations, since they are considerably
larger (by a factor of ten and more) compared to mass-loss rates from
O-stars of similar luminosity. Though \citet{LA93} showed that
line-overlap effects, coupled with a significantly stratified
ionization balance, can help a lot to increase the mass-loss, it were
\citet{GH05, GH06, GH07,GH08} who were the first to calculate
consistent WR-wind models with the observed large mass-loss rates in
parallel with high terminal velocities (2000 - 3000 \kms). They showed
that a high Eddington-$\Gamma$ is necessary to provide a low effective
gravity and to enable a deep-seated sonic point at high temperatures.
Then, a high mass-loss rate leading to an {\it optically thick} wind
can be initiated either by the `hot' Fe-opacity bump (around 160 kK,
for the case of WCs and WNEs) or the cooler one (around 40 to 70 kK,
for the case of WNLs)\footnote{The importance of these opacity bumps
had already been pointed out by \citet{NL02}.}. The high initiated 
mass-loss rates can then be further accelerated by efficient multi-line 
scattering in a stratified ionization balance (see above), at least if 
the outer wind is significantly clumped. 

Alternative wind models for Very Massive Stars in the range of $40~\msun 
< M < 300~\msun$ (i.e., including models which should display WR
spectral characteristics) have been constructed by \citet{Vink11} (but 
see also \citealt{Pauldrach12}), who argue that for $\Gamma_{\rm e} > 0.7$
these line-driven winds become optically thick already at the sonic
point, which enables a high \mdot\ with a steeper dependence on
$\Gamma_{\rm e}$ than for optically thin winds\footnote{The actual
origin of this behavior is still unclear, but the authors of
this review speculate about a higher efficiency of multi-line
effects.}. Recently, \citet{Bestenlehner14} investigated the mass-loss
properties of a sample of 62 O, Of, Of/WN, and WNh stars within the
Tarantula nebula, observed within the VLT FLAMES Tarantula Survey
\citep{Evans11} and other campaigns. Indeed, they found a change in
the slope of $d\log \mdot/d\log \Gamma_{\rm e}$ towards higher values.
However, this change occurs already at $\Gamma_{\rm e} = 0.25$, i.e.,
(much) earlier than predicted by \citet{Vink11}, and more consistent
with the models by \citet{GH08}.
Moreover, at least for Of and Of/WN stars there is still the
possibility that the conventional CAK theory (which already includes a
tight dependence on $\Gamma_{\rm e}$, cf. Eq.~\ref{mdotscal}) remains
applicable, though with a lower $\alpha$ (0.53 instead of 0.63) than
for typical O-stars. Thus, a number of issues still need to be worked
out before these optically thick winds are fully understood.

\section{Winds from Red Supergiants} 

Typical mass-loss rates from RSGs\footnote{for structure and stellar
parameters, see Wittkowski, this Volume \hfill \quad} range from $10^{-5}$ to
$10^{-4}$ \msunyr, with terminal velocities on the order of 20 to 30
\kms\ ($\approx \vesc/3$). Note that the atmospheres of RSGs consist
of giant convective cells, with diameters scaling with the vertical
pressure scale height \citep{SteinNordlund98, Nordlund09}. 
%($\approx$ 10 \rsun, for a radius of many hundreds of \rsun). 
Though the physics of RSG winds is still unknown,
a similarity to the dust-driven winds from (carbon-rich = C-type)
AGB-stars is often hypothesized\footnote{According to
\citet{Hoefner08}, dust-driving might be also possible in oxygen-rich
(M-type) AGB-atmospheres, if prevailing conditions allow forsterite
grains (Fe-free olivine-type, Mg$_2$SiO$_4$) to grow to sizes in the
micro-meter range.}. In these stars, stellar (p-mode) pulsations or
large scale convective motions lead to the formation of outward
propagating shock waves, that lift the gas above the stellar surface,
intermittently creating dense, cool layers where dust may form.
Similar to line-driven winds, these
dust grains are radiatively accelerated, and drag the gas via
collisions (for a review, see \citealt{Hoefner09}). 

\citet{Josselin07} alluded to some problems of this scenario when
applied to RSG-winds. Namely, RSGs have only irregular,
small-amplitude variations, which makes lifting the gas difficult in 
the first place, and moreover the dust seems to form much further out
than in AGB-stars. On the other hand, they also pointed out that
turbulent pressure related to convection helps in lowering the {\it
effective} gravity\footnote{As a side note, we might ask whether the
well-known mass-discrepancy for O-type dwarfs might be related to the
neglect of a potentially large turbulent pressure in present
atmospheric models (see Markova \& Puls, this Volume).} and thus the
effective escape velocity, $g_{\rm eff} \approx g/(1+\mu v^2_{\rm
turb}/(2k_{\rm B}T))$, with $\mu$ the mean molecular weight, and
suggested that radiative acceleration provided by molecular lines
might help lift the material to radii where dust can form, though
without any quantitative estimate. To conclude, further 
investigations and simulations to explain RSG-winds are urgently
needed.

\section{Continuum-driven winds\label{porwinds}} 

As a prelude to the following scenario, let us check in how far a 
hot stellar wind can be also driven by pure continuum processes, with
major opacities due to bound-free absorption and Thomson scattering. 

\smallskip
\noindent
{\bf The simple picture.} Since these opacities
(per volume) scale mostly with linear density, the corresponding
mass absorption coefficients (frequential and flux-weighted, see
Eq.~\ref{gradformal}) do not display any explicit density dependence. 
Consequently, the {\it total} Eddington-Gamma,
\beq
\Gamma_{\rm tot}(r)=\frac{g_{\rm rad}(r)}{g_{\rm grav}(r)} = 
\frac{\kappa_{\rm F}(r) L}{4\pi c G M} \rightarrow \Gamma_{\rm cont}(r)
\eeq
is density-independent as well (contrasted, e.g., to the case of
line-driving), and it seems that basically {\it any}
\mdot\ might be accelerated\footnote{since the equation of motion does
no longer depend on $\rho$ \hfill \quad} as long as $\Gamma_{\rm cont}(r)$
increases through the sonic point, with $\Gamma_{\rm cont}(r_{\rm s})
= 1$, and remains beyond unity above. 

As pointed out by \citet{OG97}, however, photon tiring
(Sect.~\ref{basics}) decreases the available luminosity, $L(r) <
\Lstar$, and thus the mass-loss rate is still restricted by
$\mdot \le \mdot_{\rm tir}$. Moreover, the complete process requires a
substantial fine-tuning to reach and {\it maintain} $\Gamma_{\rm cont}
\ge 1$ in (super-) sonic regions, and such a `simple' continuum-driving 
is rather difficult to realize.

\smallskip
\noindent
{\bf Super-Eddington winds moderated by porosity.} 
Whilst, during `quiet' phases, LBVs lose mass most likely via ordinary
line-driving (cf. Sect.~\ref{ldw}), they are also subject to one or
more phases of much stronger mass loss. E.g., the giant eruption of
$\eta$ Car with a cumulative loss of $\sim$10 \msun\ between 1840 and
1860 \citep{Smith03} corresponds to $\mdot \approx$~0.1-0.5~\msunyr,
which is a factor of 1000 larger than that expected from a line-driven
wind at that luminosity. Such strong mass loss has been frequently
attributed to a star approaching or even exceeding the Eddington
limit. 

Building upon pioneering work by \citet{Shaviv98, Shaviv00,
Shaviv01b}, \citet{OGS04} developed a theory of ``porosity-moderated''
continuum driving in such stars, where
the dominating acceleration is still due to continuum-driving, mostly
due to electron scattering, i.e., $\Gamma_{\rm tot} \rightarrow \Gamma_{\rm
cont} \approx \Gamma_{\rm e} >1$. 

For stars near or (formally) above the Eddington limit, non-radial instabilities
will inevitably arise and make their atmospheres inhomogeneous
(clumpy), see, e.g., \citet{Shaviv01a}. As noted by \citet{Shaviv98}, the
{\it porosity} of such a structured medium can reduce the radiation
acceleration significantly (photons `avoid' regions of enhanced density), by
lowering the {\it effective} opacity in deeper layers, $\Gamma^{\rm
eff}_{\rm cont} < 1$ for $r < r_{\rm s}$, thus enabling a
quasi-hydrostatic photosphere, but allowing for a transition to a
supersonic outflow when the over-dense regions become optically thin
due to expansion, $\Gamma^{\rm eff}_{\rm cont} \rarrow \Gamma_{\rm
cont} > 1$ for $r > r_{\rm s}$.

The effective opacity in a porous medium consisting of an
ensemble of clumps can be derived from rather simple arguments
\citep{OGS04}, but here it is sufficient to note that for
{\it optically thick clumps} and $\rho$-dependent opacities  
\beq
\kappa_{\rm F}^{\rm eff}(r)=\frac{1}{h\langle\rho\rangle(r)} \ll
\kappa_{\rm F}(r),
\eeq
the {\it effective} opacity (here: the effective mass absorption
coefficient) becomes grey and much smaller than the original one. In
this equation, $\langle\rho\rangle(r)$ is the {\it mean} density of
the medium, and $h$ the so-called porosity length, which is the
photon's mean free path for a medium consisting of optically thick
clumps. Thus, $\kappa_{\rm F}^{\rm eff}(r)$ and consequently
$\Gamma_{\rm cont}^{\rm eff}(r)$ have a specific density dependence
around the sonic point ($\propto \langle\rho\rangle^{-1}$), and there
is a corresponding, well-defined \mdot\ which can be initiated and
accelerated.

If one now considers clumps with a range of optical depths, 
distributed according to an exponentially truncated 
power-law with index $\alpha_{\rm p}$ \citep{OGS04}, one obtains 
for sound speed $a$ and pressure scale height $H$,
\beq
\mdot (\alpha_{\rm p} = 2) = \bigl(1-\frac{1}{\Gamma_{\rm cont}}\bigr)
\frac{H}{h}\frac{\Lstar}{a c} = \frac{\mdot (\alpha_{\rm p} = 1/2)}
{4 \Gamma_{\rm cont}},
\eeq
where the mass-loss rate for a canonical $\alpha_{\rm p} = 2$ model 
(obtained from a clump-ensemble that follows Markovian statistics, 
\citealt{Sundqvist12b, Owocki14}) saturates for very high $\Gamma_{\rm
cont}$, but where the alternative $\alpha_{\rm p} = 1/2$ model can
give an even higher mass loss for such cases, approaching the tiring
limit under certain circumstances, and being on order the mass loss
implied by the ejecta of $\eta$~Car for an assumed $h \approx H$
\citep{OGS04, Owocki14}. Detailed simulations are needed here to
further constrain the clump-distribution function and the porosity 
length in these models.

%following a power-law with exponent $(\alpha_{\rm p}-1)$,
%mass-loss rates as implied from the ejecta of $\eta$~Car can be
%obtained, if $\alpha_p < 1$. For $\Gamma_{\rm cont} > 3{\ldots}4$,
%
%\beq
%\mdot \approx \Bigl[\bigl(1-\frac{1}{\Gamma^2_{\rm cont}}\bigr)
%\frac{H}{h}\frac{\Lstar}{a c}\Bigr] \Gamma_{\rm cont}^{1/\alpha_{\rm
%p}-1},
%\eeq
%
%\mdot\ becomes enhanced by a factor increasing with
%$\Gamma^{1/\alpha_p -1}$, compared to a porous model with a single
%clump optical depth (the expression in square brackets), approaching
%the tiring limit under certain circumstances. Here, $a$ is the sound
%speed, $H$ the pressure scale height, and it is assumed that $h
%\approx H$.

Nonetheless, together with quite fast outflow speeds, \vinf\ $\approx
\mathcal{O}(\vesc)$, and a velocity law corresponding to $\beta =1$,
the derived wind structure in such a porosity-moderated 
wind model may actually explain the observational constraints of 
giant outbursts in $\eta$~Car and other LBVs. Moreover,
the porosity model retains the essential scalings with gravity and
radiative flux (the von Zeipel theorem, cf. Sect.~\ref{rot}) that
would give a rapidly rotating, gravity-darkened star an enhanced polar
mass loss and flow speed, similar to the bipolar Homunculus nebula.
Note that continuum driving (if mostly due to Thomson scattering) does
{\it not} require the presence of metals in the stellar atmosphere.
{\it Thus, it is well-suited as a driving agent in the winds of
low-metallicity and First Stars, and may play a crucial role in their
evolution.}

\section{Additional physics in line-driven winds}
\label{addeffects}

In this last section we now return to line-driven winds, and
discuss specific conditions and effects which might influence their appearance.

\subsection{Rapid rotation\label{rot}}

When stars rotate rapidly, their photospheres become oblate (because
of the centrifugal forces, see \citealt{Collins63, Collins66}), the
effective temperature decreases from pole towards equator (`gravity
darkening', \citealt{Zeipel24, MaederIV}), and the winds from typical
O-stars are predicted to become {\it prolate} (because
of the larger illuminating polar fluxes), with a fast and dense polar
outflow, and a slow and thinner equatorial one
\citep{CO95}\footnote{All these effects become significant if the
rotational speed exceeds roughly 70\% of the critical one. Note also
that cooler winds might retain an oblate structure, if the ionization
balance decreases strongly from pole to equator, and the effective
number of driving line increases in parallel.}. 

Whilst the basic effects of stellar oblateness and gravity darkening 
have been confirmed by means of interferometry \citep[see
also van Belle, Meilland, Faes, this Volume]{deSouza03, Monier07}, the
predictions on the wind-structure of rapidly rotating stars have {\it
not} been verified by observations so far (\citealt{Puls11} and
references therein): first, only few stars in phases with extreme
rotation are known (but they exist, e.g., \citealt{Dufton11}), and
second, the tools to analyze the atmospheres and winds (multi-D
models!) of such stars are rare. Even though \citet{Vink09b} (see also Vink,
this Volume) reported, analyzing data obtained by means of
linear \Ha\ spectro-polarimetry, that most winds from rapidly rotating
O-stars are spherically symmetric (actually, they looked for disks),
the asymmetry predicted for the winds of this specific sample should
be rather low anyway.

Besides the polar-angle dependence of \mdot\ induced by rotation, also
the global mass-loss rate becomes modified; a significant increase,
however, is only found for rapid rotation {\it and} a large $\Gamma_{\rm
e}$, with a {\it formal} divergence of \mdot\ -- which at least needs to be
corrected for photon tiring effects -- at the so-called
$\Omega\Gamma$-limit. For details, see \citet{MM00}. 

Finally, for near-critically rotating stars, mass loss might also occur
via decretion disks \citep{Krticka11}. The corresponding \mdot\ from such 
decretion disks can be significantly {\it less} than the spherical, wind-like mass loss (aka
`mechanical winds') previously assumed in evolutionary calculations.

\subsection{Magnetic fields\label{mag}}

Recent spectropolarimetric surveys (mostly performed by the
international Magnetism in Massive Stars, MiMeS, collaboration, e.g.,
\citealt{Wade12}, and work done by S. Hubrig and collaborators, e.g.,
\citealt{Hubrig13}; see also Grunhut, Morrell, this Volume) have
revealed that roughly 10\% of all massive stars have a large-scale,
organized magnetic field in their outer stellar layers, on the order
of a couple of hundred to several thousand Gauss. The origin of these
fields is still unknown, though most evidence points to quite stable
fossil fields formed sometimes during early phases of stellar
formation \citep{Alecian13}. The interaction of these fields with a
line-driven stellar wind has been investigated by ud-Doula, Owocki and
co-workers in a series of publications (summarized in
\citealt{udDoula13}, see also ud-Doula, this Volume). In the
following, we concentrate on {\it slowly} rotating magnetic O-stars
(spectroscopically classified as O\,f?p, \citealt{Walborn72}), which
give rise to so-called `dynamic magnetospheres' \citep{Sundqvist12}.

The most important quantity to estimate the influence of a magnetic
field on the wind is the ratio of magnetic to wind energy, 
\beq
\eta=\frac{B^2/8\pi}{\rho v^2/2} =\frac{B_\ast^2 \Rstar^2}{\mdot
\vinf^2}f(r):=\eta_\ast f(r),
\eeq
with $\eta_\ast$ the so-called confinement parameter. E.g., for a
typical O-supergiant a B-field of 300 Gauss is required to reach
$\eta_\ast$ = 1 (for $\eta_\ast < 1$, the wind is not much disturbed).
In the case of an $\eta_\ast$ significantly greater than unity, 
to a good approximation the corresponding Alv\'en radius (which is
the maximum radius for closed loops, and determines whether the wind
is confined in such loops), can be expressed 
by $R_{\rm A} \approx \Rstar \eta_\ast^{1/4}$. 

An instructive example for a strongly confined wind can be found in,
e.g., \citet{Sundqvist12}, who performed hydro-dynamical simulations 
and H$\alpha$ radiative transfer calculations for the prototypical 
Of?p star HD\,191612, with a B-field of $\approx$ 2,500 Gauss
corresponding to $\eta_\ast=50$. In this model, the field loops are
closed near the equatorial plane ($R_{\rm A} \approx 2.7 \Rstar$), and
the confined wind is accelerated and channeled upwards from
foot-points of opposite polarity. The flows collide near the loop
tops, forming strong shocks with hard X-ray emission \citep{Gagne05}. 
The shocked, very dense material then cools and becomes accelerated
{\it inwards} by the gravitational pull, emitting strongly in, e.g.,
the optical \Ha\ line. Whilst there are complex infall patterns along
the loop lines, the field lines in polar regions are still open and
the polar wind remains almost undisturbed.  The {\it infalling}
material of dynamical magnetospheres reduces the global mass-loss
rates, for large $\eta_\ast$ by a factor $\sim$\,5 compared to
non-magnetic winds \citep[analytic scaling relations
available]{udDoula08}. Note also that the non-spherical structures
require well-suited diagnostic methods.

\subsection{Inhomogeneous winds -- a few comments\label{inhom}} 

We finish this review with few comments about the presence and impact
of small-scale wind inhomogeneities (for a detailed discussion and references,
also regarding large-scale inhomogeneities, see, e.g., \citealt{puls08b}).

Over the last two decades, a multitude of direct and indirect
indications has been accumulated that hot star winds are inhomogeneous
on small scales, i.e., consist of over-dense (compared to the
mean-density) clumps and an inter-clump material which is frequently
assumed to be void (but see \citealt{Surlan13, Sundqvist14}).

The most likely origin is the line-driven (or line-deshadowing)
instability, which, for short-wavelength perturbations, can be
summarized by $\delta g_{\rm rad}^{\rm lines} \propto \delta v$,
giving rise to strong, outward propagating reverse shocks emitting in
the X-ray regime, and a wind-structure consisting of fast and thin
material (inter-clump matter), and dense, spatially narrow 
clumps moving roughly at the speed of smooth-wind models.

In dependence of the considered absorption process and wavelength,
clumps can be optically thin (`micro-clumping') or optically thick
(`macro-clumping'/ porosity, see Sect.~\ref{porwinds}). Moreover, line
processes are prone to porosity in velocity space. To account for the
effects of wind-inhomogeneities, a simplified treatment is typically 
employed (both within radiative transfer and when calculating the occupation
numbers), based on a one- or two-component description, with parameterized
clumping properties such as volume-filling factors, over-densities,
etc.

The major impact of these inhomogeneities regards the various
mass-loss diagnostics. If clumps are optically thin for
$\rho^2$-dependent opacities (e.g., \Ha\ and IR diagnostics in
O-stars), the actual rates turn out to be lower than derived from
smooth-wind models. If clumps are optically thick and/or velocity
porosity needs to be accounted for (e.g., UV-resonance lines), the
final rates are larger than derived from models assuming optically
thin clumps alone. On the other hand, mass-loss diagnostics based on
bf-absorption (by the cool wind) of X-ray line emission from the above
wind-embedded shocks is particularly robust, since it remains
uncontaminated by inhomogeneities in typical O-star winds: first,
there is no direct effect from micro-clumping, because the involved
bf-opacities (per volume) scale with $\rho$, and second,
porosity-effects are negligible or low (e.g., \citealt{Cohen10,
Cohen13, Leutenegger13, Herve13}). 

Comparing now `observed' O-star mass-loss rates with theoretical ones
(from \citealt{vink00}) used in stellar evolution, there is the
following status quo\footnote{to, e.g., clarify corresponding
statements in the recent review by \citet{Smith14} that are somewhat
simplified in this respect.}: These theoretical mass-loss rates are
(i) a factor of 2-3 {\it lower} than those from standard \Ha\
diagnostics assuming a smooth wind; (ii) roughly consistent with radio
mass-loss rates assuming a smooth wind; (iii) a factor of 2-3 {\it
larger} than recent diagnostics of Galactic O-stars accounting
adequately for wind inhomogeneities (\citealt{Najarro11}: mostly IR-lines; 
%\citealt{Bouret12}: UV/FUV; 
\citealt{Cohen14}: absorbed X-ray line emission;
\citealt{Sundqvist11}, \citealt{Surlan13}, \citealt{Sundqvist14}:
UV-lines including velocity porosity+optical lines).

Of course, further investigations and larger samples are certainly 
required to prove this discrepancy, but particularly the X-ray results
are a strong argument. Indeed, there are various possibilities for a
potential overestimate of theoretical mass-loss rates, summarized by
\citet{Sundqvist13}. A rather promising explanation relates to (so far
neglected) effects from velocity porosity when calculating the line
force, which can lead to reduced theoretical mass-loss rates if
already present in the lower wind (\citealt{Sundqvist14}, see also
\citealt{Muijres11}). 

In summary, there is still much to do, and the physics of massive
star winds remains a fascinating topic!

\acknowledgements{JOS and JP gratefully acknowledge support by the
German DFG, under grant PU117/8-1. A travel grant by the University of
Geneva is gratefully acknowledged by NM and JP.}  

\bibliographystyle{iau307}
\bibliography{MyBiblio}

\begin{discussion}

\discuss{de Koter}{We find that the line-driven winds of O stars at
metallicities below that of the SMC do not seem to obey the theory of
line driving: the mass-loss rates are higher (Tramper et al. 2011).
What are your ideas on an explanation of this peculiar behaviour?}
\discuss{Puls}{Actually, this is not completely clear. As will be
shown in the next talk by M. Garc\'ia, there might be a bias on
\mdot\ due to variations in the ratio of \vinf/\vesc,
and in this particular case the iron abundance (driving agent) might
be higher than implied by the oxygen abundance.}
\discuss{de Koter}{The winds of RSG are very difficult to understand,
so currently we focus on understanding AGB winds. Though ideas have
been put forward to explain the O-rich outflows, in my view there are
still fundamental problems. These winds can only be driven through
scattering on large ($0.3\,\mu\mathrm{m}$) grains and it is not clear
at present how to grow such large grains in the warm molecular layer.}
\discuss{Puls}{Completely agreed.}
\discuss{Khalak}{Can you explain the reasons for the excitation of
atoms that increases line opacity and causes optically thick wind
having $\Gamma_\mathrm{e}>0.7$ in WR stars?}
\discuss{Puls}{1. Lines become more easily optically thick because of
the higher density. Line overlap effects are particularly effective
for optically thick lines. 2. Due to the higher wind density, the
ionisation/excitation couples closer to the local electron
temperature. Thus there is an ionisation stratification (which again
allows for efficient line overlap), and the occupation numbers of the
excited levels increases as well (closer to Boltzmann).}
\discuss{Weis}{You showed how the giant eruption could be explained.
Can that model also explain the bipolar structure?}
\discuss{Puls}{Indeed, the radiative acceleration due to
porosity-moderated continuum driving has a similar dependence on polar
angle as in line-driven winds, for rapidly rotating stars/winds. Thus,
also here a prolate structure is expected.}
\discuss{Noels}{Would it be possible to have a table with simple
formulas of the mass-loss rates across the upper part of HR diagram,
with the uncertainties (even large!)? This could appear ``as of
today'' in this proceeding.}
\discuss{Puls}{I will try my best, but only scaling relations can be
provided in some cases (e.g., excluding RSGs). The absolute numbers of
mass-loss rates are heavily debated, e.g. due to the impact of wind
inhomogeneities. {\it Added after review has been finished:} Actally,
such a table could not be provided, given the limited space and the
state of our knowledge. Anyhow, all relevant references
have been provided, but there is still strong disagreement on the
uncertainties. And since factors of two {\it are} important, quoting
disputed values with large error bars is not meaningful.}

\end{discussion}

\end{document}